# Stability of glycol nanofluids – the consensus between theory and measurement

Ibrahim Palabiyik, Sanjeeva Witharana*, Zenfira Musina, Yulong Ding

*Institute of Particle science and Engineering, School of Process, Environmental and Materials Engineering, University of Leeds,*
*Woodhouse Lane, LS2 9JT Leeds, UK*
*Tel: +441133432543, Fax: +441133432384*

*Corresponding author: switharana@ieee.org

**Abstract:** Formulation of stable nanofluids containing ZnO, $Al_2O_3$ and $TiO_2$ nanoparticles in propylene glycol (PG), ethylene glycol (EG) and 50wt% mixtures of PG and EG in water (WPG, WEG) were investigated, with and without the presence of surfactants. Nanofluid samples of particle concentrations 1-9wt% were prepared by dispersive method. Surfactant presence was in the range of 0-1wt%/wt% of nanoparticles. Visual observation, particle size measurement and zeta potential analysis were performed to evaluate the dispersion stability. In overall the PG-based samples were found to be the most stable suspensions. The effect of base fluid on particle size and the effect of day light on nanofluid stability were also examined as a function of time. $TiO_2$-PG samples showed a colour change when exposed to sunlight. Sunlight also caused the PG based $TiO_2$ and $Al_2O_3$ nanofluid to increase their particle sizes by up to 45% in the course of 3 days. As for stability, the sedimentation velocity was observed to be a key parameter. Finally by comparison of settling theory with experiments, a stability boundary was demarcated to identify stable and unstable nanofluids.

## 1. INTRODUCTION

Nanofluids, which are referred to as dilute liquid suspensions of nanoparticles in common fluids, have been a subject of great interest in the past decade due to their unique thermophysical properties and heat transfer behaviour. Experiments have shown that nanofluids were able to enhance the thermal conductivity and convective heat transfer by large margins [1-3], and critical heat flux by up to 300% [4, 5]. In many instances, nanofluids also enhanced the pool boiling heat transfer [6].

Dispersion method, also called two-step method, is generally favoured for preparing nanofluids containing high volume fraction metals, oxides and carbon nanotubes. Here the dry nanopowder is dispersed in the liquid by application of one or many dispersion techniques [7]. This method is more economical in comparison to one-step method, due to the low cots of nanopowders in the market. Decline of the price of nanopowders is a result of the rapid development of high-throughput nanoparticle production technologies over the years. Nanofluids prepared by dispersion method however commonly have shown a stability problem [8-10]. These nanofluids gradually start to settle after a period of time depending on the properties of base liquid, surfactant or dispersant used, type of nanoparticles, and the likelihood of nanoparticles to aggregate. The validity of a nanofluid is as only long as it is stable. An agglomerated nanofluid is different in properties, and may cause operational problems similar to those encountered with micron-sized particulate suspensions; sedimentation and clogging of the system. Unstable nanofluids moreover are most likely a root cause for the wide discrepancies in literature data on their heat transfer behaviour.

Therefore, the preparation of stable nanofluids is undoubtedly the first step in nanofluid research and applications. Colloids theory states that there is a critical radius below which the sedimentation of a particle ceases due to counterbalancing of gravity forces by the Brownian diffusion. Keeping the size of nanoparticles sufficiently small in the liquid should therefore be the focal point in the formulation exercise. On the other hand, smaller nanoparticles possess higher surface energies that causes higher tendency to build agglomerates among them. Furthermore, tiny particles causes higher electrolyte concentration in the nano-suspensions; the reason is large surface area contains large amount of ionisable sites. In relation to this Jailani et al. [11] observed that high electrolyte concentration in nanofluids causes decrease in zeta potential. Keeping the particle sizes very small can hence be counterproductive for a stable nanofluid. The challenge of formulating stable nanofluids is to prevent coalescence of nanoparticles while keeping their size and concentration optimum in the base liquid.

Ultrasonic agitation and/or mechanical stirring are the widely used techniques for breaking the large agglomerates into smaller pieces and to ensure good dispersion of particles in the liquid [12]. Changing the pH of suspensions and adding surfactants can prevent them from coalescence [13-15]. In some works, surfactants have been avoided as they are believed to influence the thermophysical properties of the suspension [12, 16] and owing to the risk of their failure at high temperatures.

The most simple, reliable and widely used technique to evaluate the stability of a nanofluid is the sedimentation method, also known as the settling bed [17]. Either the light absorbance is measured or the bed height is visually monitored over a certain period of time. However, the main drawback of using this method on fairly slow settling suspension is the long observation times. A faster technique is the time-resolved measurement of the zeta potential of the sample. Down side of this method is, it imposes restrictions on the viscosity and particle concentration of the samples.

Despite the fact that the formulation of stable nanofluids is the foundation for clean and consistent measurements, a systematic examination on this topic is yet to be seen in literature. Besides there are very few accounts on propylene glycol based nanofluids [18]. This paper addresses those two aspects. $Al_2O_3$ and $TiO_2$ were chosen considering the huge interest in them in heat transfer applications [19, 20], and ZnO was chosen for their prominence in anti-microbial behaviour [21].

## 2. EXPERIMENTS

### 2.1 Materials

Hydrophilic spherical nanoparticles of $TiO_2$ and $Al_2O_3$ (supplied by Degussa, Germany) and ZnO (from NanoTek and Alfa Aesar) were purchased in the forms of dry powder. Physical properties of these nanoparticles are given on Table 1. Ethylene glycol (99% purity from AlfaAesar) and propylene glycol (98% purity from Fluka Analytics) were used as dispersants without further purification. Distilled water was used to prepare 50-50wt% mixtures, viz., water-ethylene glycol (WEG) and water-propylene glycol (WPG). pH adjustment of nanofluids was achieved through analytical grade 0.1M NaOH and HCl.

Table 1: Physical properties of nanoparticles

| Nanoparticle material | Primary size (nm) | Specific surface area ($m^2/g$) |
|---|---|---|
| $TiO_2$ | 21 | 50±15 |
| $Al_2O_3$ | 13 | 100±15 |
| ZnO | 40-100 | 10-25 |

Chemical compositions of the surfactants used in this study are given in Table 2.

Table 2: A description of surfactants

| Surfactant | Supplier | Chemical composition |
|---|---|---|
| Disponil A 1580 | Cognis | Mixture of ethoxylated linear fatty alcohols |
| Hydropalat 5040 | Cognis | Aqueous solution of sodium polyacrylate |
| Antiterra 250 | BYK | Solution of an alkylolammonium salt of a high molecular weight acidic polymer |
| Disperbyk 190 | BYK | Solution of a high molecular weight block copolymer with pigment affinic groups |
| Hypermer LP1 | Croda | Polycondensed fatty acid |
| Aerosol TR-70 | Cytec | Sodium bistridecyl sulfosuccinate (anionic 70% solution in ethanol and water) |
| Aerosol TR-70 HG | Cytec | Sodium bistridecyl sulfosuccinate (anionic 70% solution in hexylene glycol and water) |
| Aerosol OT-70 PG | Cytec | Sodium dioctyl sulfosuccinate (anionic 70% solution in propylene glycol and water) |
| Gum Arabic | MP Biomedicals | Natural polysaccharides and glycoproteins complex |

**2.2 Dispersion stability evaluation**

In this work care was taken to avoid wet milling as a part of formulation sequence. To break agglomerates and reduce the particle size, long term ultrasonication was applied at 37 kHz. The instrument used for this purpose was a Digital Sonicator (Model S70H from Elma, Germany). Dispersion characteristics of suspensions were evaluated by visual inspection, particle size measurements, and zeta potential analysis. Particle size and zeta potential measurements were conducted using a Zetasizer Nano ZS device (from Malvern Instruments) equipped with a MPT-2 autotitrator. Tititrations were performed at 20°C temperature and Smoluchowski model embedded in the device. Thicknesses of deposits at the bottom of containers (vials) were measured one month after sonication.

**3. RESULTS AND DISCUSSIONS**

Under this section $TiO_2$, $Al_2O_3$, and ZnO nanofluids will be treated separately.

*3.1 $TiO_2$ Nanofluids*

$TiO_2$ samples contained a wide range of particle concentrations from 1-9wt%. Table 3 shows the visual observations after 2 months from formulation. $TiO_2$ –PG and $TiO_2$ –EG nanofluids were found to have deposits at the bottom of vials except the 1wt % PG-$TiO_2$ sample. Deposited particles are thought to be of big agglomerates which could not be broken down into small pieces by applying sonication. In order to get rid of these big agglomerates, the visually-stable upper parts of the samples were decanted into new vials leaving behind the sediments. Out of these, only the PG based nanofluids remained stable after 2 months.

Table 3: Observations for $TiO_2$ nanofluids

| Wt% | Base fluid | pH | Son. Time (hr) | Observation |
|---|---|---|---|---|
| 1 | PG | Not adjusted | 38 | Stable |
| 6 | PG | " | 38 | Stable after decant |
| 9 | PG | " | 38 | Stable after decant |
| 1 | EG | " | 38 | Thin complete sediment |
| 6 | EG | " | 38 | 1 mm sediment |
| 9 | EG | " | 38 | 2 mm sediment |
| 1 | WPG | Not adjusted | 24 | Very little sediment |
| 1 | WPG | 3 | 24 | Very little sediment |
| 1 | WPG | 7 | 24 | Phases separated |
| 1 | WPG | 9 | 24 | Very little sediment |
| 1 | WPG | 11 | 24 | Very little sediment |
| 1 | WEG | Not adjusted | 24 | Very little sediment |
| 1 | WEG | 3 | 24 | Phases separated |
| 1 | WEG | 7 | 24 | Phases separated |
| 1 | WEG | 9 | 24 | Very little sediment |
| 1 | WEG | 11 | 24 | 1 mm sediment |

PG based nanofluids showed good overall stability for all different concentrations. For $TiO_2$ nanoparticles therefore PG was observed to be better base liquid than EG in terms of stability. Likewise, less amount of sediment was observed in $TiO_2$–WPG samples than $TiO_2$–WEG samples. One possible explanation is the viscosity of these liquids; viscosity of PG is 52mPa.s at 20°C, while for EG it is around 16mPa.s. Furthermore, increase in nanoparticle concentration always caused increase of supporting the concept that more dense suspensions increase the rate of re-agglomeration of nanoparticles.

Previously Chen et al. [7] had formulated stable $TiO_2$–EG nanofluids with up to 8wt% particle concentration. In this work however the stability was weaker despite the same nanoparticle source and formulation procedure were followed. The difference may be attributed to the methods of handling and partial agglomeration of particles due to aging of the nanopowder.

Titrations were performed for WEG and WPG based 1wt% $TiO_2$ samples in order to find their iso-electric points (IEP). As of figure 1, the $TiO_2$–WEG nanofluids had a zeta potential close to -40mV, which is adequate for a fairly table suspension, in the pH range of 6.2-7.8. Recall that the initial pH of $TiO_2$–WEG nanofluid was 6.8. Hence this sample is expected to be stable even without pH adjustment. Closer to its IEP (pH 4.7), this nanofluid should be unstable.

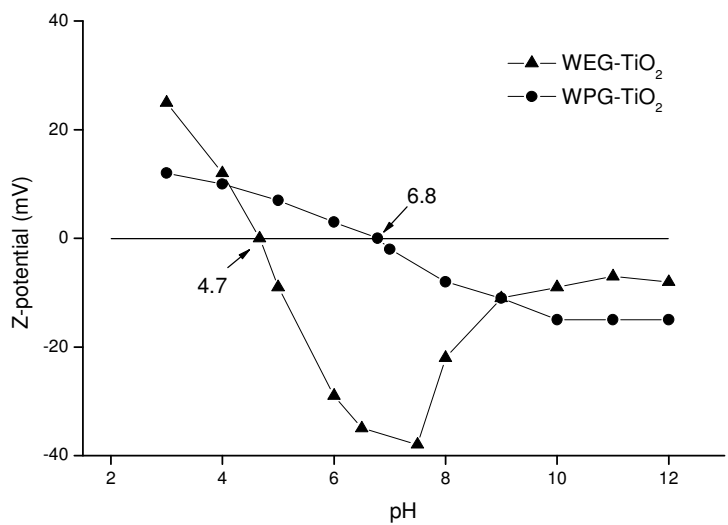

Figure 1: Zeta potential curves for 1wt% suspension of $TiO_2$ in WEG and WPG at 20°C

On the other hand, zeta potential of $TiO_2$–WPG nanofluid fell in a relatively narrow range. Its stability may not be achieved by pH adjustment alone. Initial pH of this suspension was 6.0, which is so close to its IEP and is prone to fail immediately.

To examine the compliance of above analysis in practice, visual experiments were planned. WEG and WPG based 1wt% $TiO_2$ nanofluids were prepared and subjected to 24 hrs of sonication after dispersing the nanoparticle into the liquids. Then pH was adjusted to 3, 7, 9, and 11, and the samples were left 3 days. At the end of the waiting period, phase separation was observed in $TiO_2$–WEG sample that was at pH7. This was unusual according to titration curve in figure 1; at pH7, the zeta potential value is -39 mV. Nevertheless the $TiO_2$–WPG samples showed agreement with their titration graph. Small deposits were found in the bottoms of all 1wt% TiO$_2$ suspensions that underwent this experiment.

*Surfactant study with TiO$_2$ nanofluids*

A plausible amount of effort was dedicated to find suitable surfactants to stabilise TiO$_2$ nanofluids, with special emphasis on high temperature applications. Nine surfactants were chosen for this task as outlined in Table 4. Following Farrokhpay [22], greater trust was placed on polymeric dispersants which were claimed to be more efficient in stabilising TiO$_2$ suspensions.

The formulation procedure was as follows: The weighed amount of surfactant was mixed with the base liquid. Thereafter the nanoparticles were added to the mixture. Only exception was in the case of Aerosol surfactants, which were added after dispersion of nanoparticles in the liquid.

Out of all samples given in Table 4, the only stable suspensions were the 1wt% TiO$_2$-WEG nanofluids in the presence of Aerosol TR-70, Aerosol OT-70PG, Aerosol TR-70HG. These suspensions were generally foamy. The least amount of foam was observed in the nanofluids containing Aerosol TR-70. The supplied claimed that these surfactants were thermally stable.

Table 4: The study of surfactants

| Surfactant | Amount (wt%/wt% of nanoparticle) | Base fluid | Son. time (hr) | Result |
| --- | --- | --- | --- | --- |
| Hydropalat 5040 | 1 | WEG | 4 | 2 mm sediment |
| | 0.5 | WEG | " | Thin sediment |
| | 0.1 | WEG | " | 1 mm sediment |
| | 0.5 | WPG | " | 1 mm sediment |
| | | PG-EG | | Not soluble |
| Anti-Terra 250 | 0.5 | WEG | " | Sedimentation, foamy |
| Disperbyk-190 | 0 | all | 0 | Very foamy in all base fluid |
| Gum Arabic | 0.1 | WEG | 0.5 | Phases separated |
| | 0.05 | WEG | " | Phases separated |
| | 0.1 | WPG | " | Phases separated |
| | 0.05 | WPG | " | Phases separated |
| Disponil A 1580 | 0.1 | WEG | 4 | Foamy |
| | 0.1 | WPG | " | Foamy |
| | 0.1 | PG | " | Thin sediment |
| | 0.01 | PG | " | Thin sediment |

| | 0.5 | PG | " | Thin sediment |
| --- | --- | --- | --- | --- |
| | 0.05 | PG | " | Thin sediment |
| Hypermer LP1 | 0.1 | PG | " | Phases separated |
| Aerosol TR-70 | 0.1 | WPG | " | Foamy |
| | 0.1 | WEG | " | Stable |
| Aerosol TR-70HG | 0.1 | WPG | " | Foamy |
| | 0.1 | WEG | " | Stable, foamy |
| Aerosol OT-70PG | 0.1 | WPG | " | Thin sediment |
| | 0.1 | WEG | " | Stable, foamy |

### 3.3 Al$_2$O$_3$ Nanofluids

Table 5 present a summary of Al$_2$O$_3$ nanofluids and their visible stability after 2 months. All samples were sonicated for 16 hrs. In comparison to TiO$_2$ nanofluids, the Al$_2$O$_3$ nanofluids were more stable. This was stemming from rather high zeta potential value of Al$_2$O$_3$ nanoparticles. The zeta potential of Al$_2$O$_3$ in water was measured as 42 mV (at pH 6.5).

Table 5: Observations for Al$_2$O$_3$ nanofluids

| Wt% | Base fluid | pH | Observation after 2 months |
| --- | --- | --- | --- |
| 1 | PG | Not adjusted | Stable |
| 6 | PG | Not adjusted | Stable |
| 9 | PG | Not adjusted | Stable |
| 1 | EG | Not adjusted | Stable |
| 1 | WPG | Not adjusted | Stable for 2 weeks |
| 6 | WPG | Not adjusted | 1 mm sediment |
| 9 | WPG | Not adjusted | Very thin sediment |
| 1 | WPG | 6 | Stable |
| 1 | WEG | Not adjusted | Stable for 2 weeks |
| 6 | WEG | Not adjusted | Very thin sediment |
| 9 | WEG | Not adjusted | Very thin sediment |
| 1 | WEG | 6 | Stable |

Despite pH was not adjusted and surfactants were not added, all Al$_2$O$_3$–PG and Al$_2$O$_3$–EG samples were visually stable for 2 months. Relatively short term stability was displayed by WPG and WEG based 1wt% of Al$_2$O$_3$ samples. When their pH was raised to 6, their stability was enhanced. Raising of the particle concentration of these two nanofluids up to 6% and 9 wt% resulted in a decline of stability and creation of thin sediments just after 2 days. It is worthwhile to recall that the viscosities of WEG and WPG are far below that of EG and PG.

As was done with $TiO_2$ samples, the autotitration exercise was performed on WEG and WPG based $Al_2O_3$ nanofluids. As seen from figure 2, the $Al_2O_3$-WEG exhibited a tremendously high zeta potential ~100 mV when the pH<6. Except for the interval 8.5<pH<10.5, the nanofluid should be stable. $Al_2O_3$-WPG nanofluids yielded the maximum zeta potential ~pH6. Theoretically one can expect both WEG and WPG based $Al_2O_3$ nanofluid to show their best stability when pH<6. Now a revisit to Table 5 confirms this as a fact.

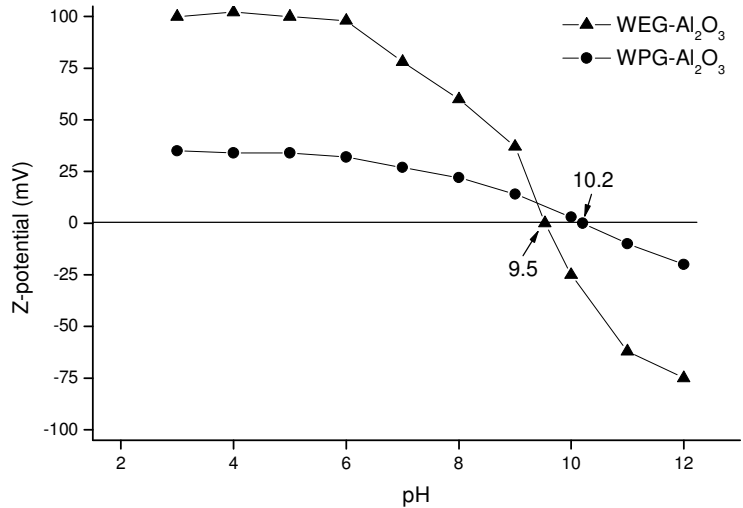

Figure 2: Zeta potential curves for 1wt% suspension of $Al_2O_3$ in WEG and WPG at 20°C

*3.3 ZnO Nanofluids*

The performance of zinc oxide suspensions was studied in different base liquids. Observations are given in Table 6. Particle size of 1 wt% ZnO in different base liquids was measured before and after 4 hrs of sonication. Figure 3 confirms that the average size of ZnO particles in PG has reduced from 360 nm to 170 nm. Note that neither surfactants were used nor pH adjusted in this formulation.

Table 6: Observations for ZnO samples after 16hrs of sonication

| Particles concentration (wt%) | Base Fluid | Z-potential (mV) | Results |
|---|---|---|---|
| 1 | PG | | Stable |
| 1 | EG | | Phase separated |
| 1 | WEG | | Phase separated |
| 1 | WPG | | 1 mm sediment |

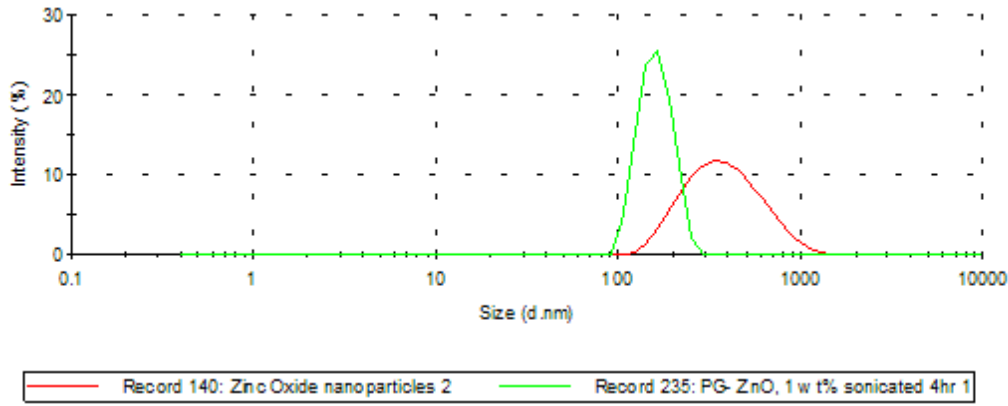

Figure 3: Initial sizes of ZnO particles in PG.

The situation of 1wt% ZnO formulations after 2 month is pictured on figure 4; from let to right are ZnO-EG, ZnO-WEG, ZnO-WPG, and ZnO-PG respectively. Only the ZnO-PG sample was observed to be stable. pH of this sample was measured as 9.

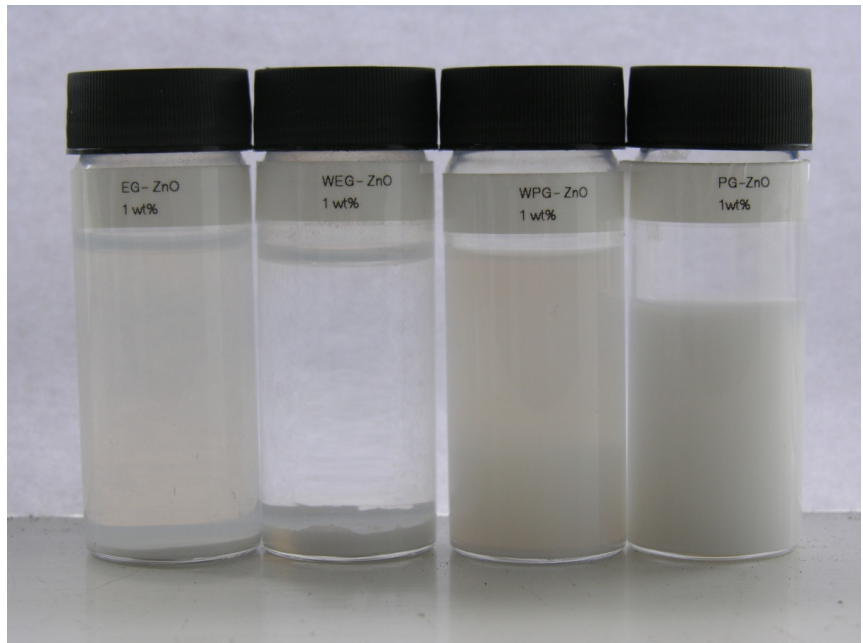

Figure 4: ZnO nanofluids 2 months after formulation

*3.4 Effect of base fluid on particle size of nanofluids*

Consider figure 5 that shows average size of $Al_2O_3$ nanoparticles in five types of base liquids including water after 12 hrs of sonication. Each line on the graph corresponds to the average of 6 consecutive measurements. The average particle size was measured as 24±4nm in EG and 84±2nm in PG, 102±5 nm in WEG, and 210±6 nm in WPG. Thus ultrasonication appears to be more effective in EG in comparison PG. It can be further deduced that the breaking of agglomerates is harder in the presence of water in base liquid. It was stated in Table 1 that the primary size of $Al_2O_3$ was claimed as 13nm by supplier. In this sense, mere 12 hrs of sonication has remarkably reduced the agglomerates in EG very close to primary particles. The present findings further suggest that the type of base liquid has a crucial role to play in breaking the bond between primary particles in agglomerates. This is a very important finding and needs to be investigated further.

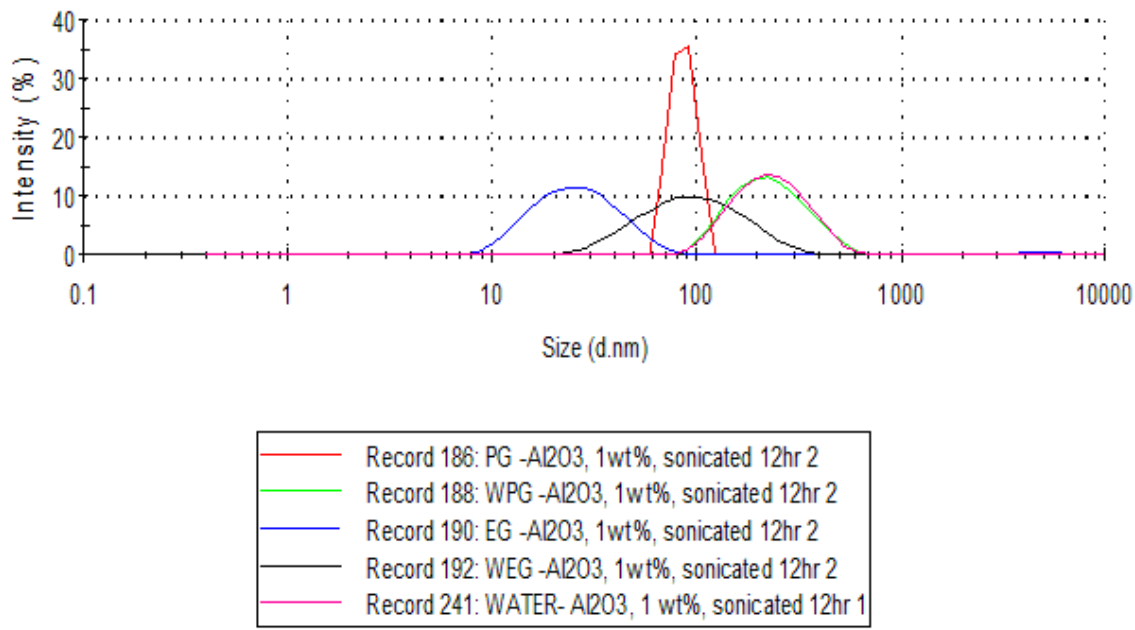

Figure 5: Average particle sizes of $Al_2O_3$ in PG, WPG, EG, WEG and water

*3.5 Effect of Day Light on Nanofluids*

According literature [23, 24], high temperatures and presence of oxygen can provoke some decomposition reactions in glycol fluids in the presence of nanoparticles. In present work, the $TiO_2$ suspensions in glycols fluids turned to blue colour after 3 weeks from formulation. It prompted to suspect that the day light may have initiated photocatalytic reactions. To investigate this further, PG based 1wt% $Al_2O_3$ and $TiO_2$ nanofluids were freshly prepared and poured into two vials. Half of them were kept in darkness and the other half on the bench top. Colour change was only observed in the TiO$_2$ containing nanofluid that was kept open to sunlight. Colour changed sample is the vial to the right hand side in figure 6. Sun light has thus triggered chemical reactions in propylene glycol fluids in the presence of TiO$_2$ nanoparticles which may have acted as a catalyst.

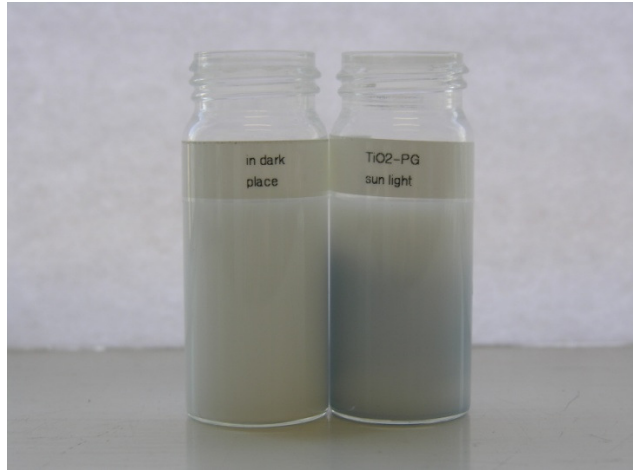

Figure 6: Effect of day light on 1wt% TiO$_2$-PG nanofluid

Average particle sizes of these four samples were then measured after 3 days of storage in darkness and open space. Zetasizer data for 1wt% Al$_2$O$_3$-PG samples are given in figure 7. The Al$_2$O$_3$ samples stored in dark place had not undergone size change. However those exposed to sunlight has increased their average size by 50nm up to ~200nm.

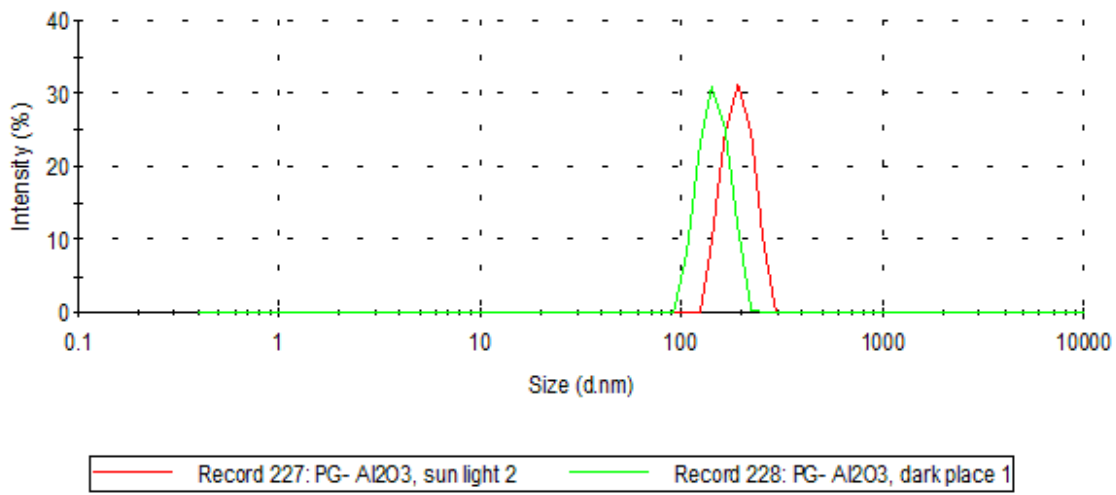

Figure 7: Effect of sunlight on particle size distribution in Al$_2$O$_3$-PG nanofluid.

In a similar manner, particle sizes in the 1wt% $TiO_2$-PG nanofluid increased from ~130 to ~180nm when exposed to sunlight but did not change in darkness.

Experiments on day light effect were continued other base liquids, viz., EG, WEG and WPG with 1wt% $TiO_2$ and $Al_2O_3$ nanoparticles. When the size measurements were taken, no difference was observed between the samples kept in darkness and samples exposed to sunlight. Therefore, it is clear that day light exposure caused agglomeration of $TiO_2$ and $Al_2O_3$ particles in PG based suspensions. Sensitivity of PG to daylight should be borne in mind when choosing storage locations for nanofluids.

### *3.6. Theoretical approach to nanofluid formulation*

According to Stokes law the sedimentation velocity (V) in a colloid can be expressed as follows;

$$V = \frac{R^2}{9\mu}(\rho_p - \rho_l).g \qquad \text{Eq (1)}$$

The rate of sedimentation decreases with decreasing particle radius (R), density difference between the particle and the liquid ($\rho_p$- $\rho_l$), and increasing base liquid viscosity (µ). These are all important parameters for a kinetically stable nanofluid. This formula was applied to the nanofluids those came under this study and plotted on figure 8. Four data columns from left to right on viscosity axis are respectively for WEG, WPG, EG and PG based nanofluids. Particle sizes for this calculation were extracted from Zetasizer measurements and given in Table 7, along with the density and viscosity of base liquids. Note that surfactants were not used in any of these samples.

Table 7: Average particle sizes in 1wt% nanofluids and, density and viscosity of liquids

|  | WEG | WPG | EG | PG |
|---|---|---|---|---|
| $TiO_2$ (nm) | 245 | 155 | 150 | 120 |
| $Al_2O_3$ (nm) | 185 | 280 | 45 | 65 |
| ZnO (nm) | 275 | 205 | 237 | 150 |
| Density (kg/m³) | 1077 | 1160 | 1130 | 1320 |
| Viscosity (mPa.s) | 0.003 | 0.006 | 0.016 | 0.052 |

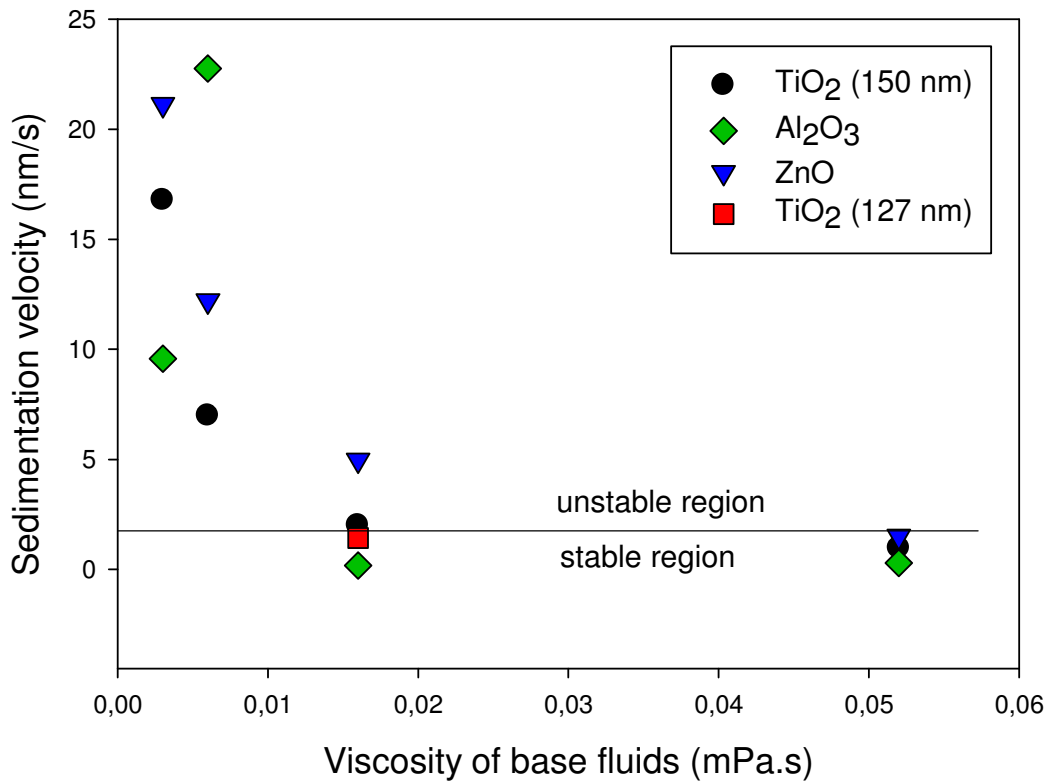

Figure 8: Stability of WEG, WPG, EG and PG based nanofluids at 20°C

At a glance on figure 8 it can be said that, all PG based nanofluids and EG based $Al_2O_3$ nanofluid are having near-zero sedimentation velocities and hence should be stable. This deduction from theory shows remarkable agreement with experimental observations. As previously stated, all PG based nanofluids and EG based $Al_2O_3$ nanofluid were visually stable for 2 months. Reason for instability of non-PG based ZnO nanofluids is the liquid viscosity. In experiments, $TiO_2$–EG nanofluids were unstable when the particle size was ~150nm. After prolonged sonication when it was reduced to ~127nm, the sample became stable for 2 months. By comparison of theory and experiment, the 2-month visual stability criterion was used to demarcate the stability boundary on figure 8.

## 4. CONCLUSIONS

A systematic and extensive study on the parameters that influence the nanofluids stability was conducted using ethylene glycol (EG), propylene glycol (PG), water- ethylene glycol 50-50wt% (WEG) and water- propylene glycol 50-50wt% (WPG) based ZnO, $Al_2O_3$ and $TiO_2$ nanofluids. Effect of particle size, surfactants, and sunlight were systematically interrogated. Finally the experimental observations were compared with Stokes predictions. The following conclusions can be derived from this exercise:

PG was found to be the leading base fluid in terms of the nanofluids stability. PG based 1wt% ZnO, $Al_2O_3$ and $TiO_2$ nanofluids exhibited excellent stability for more than 2 months since preparation. $Al_2O_3$ nanofluids in general were somewhat more stable in all types of base fluids, whereas ZnO and $TiO_2$ were stable only in PG. However Aerosol TR-70% surfactant successfully stabilised the 1wt% $TiO_2$ WEG.

Daylight was proved to cause colour change in all glycols based $TiO_2$ suspensions. Daylight also cause particle agglomeration in PG based $Al_2O_3$ and $TiO_2$ samples.

The effect of sonication on particle size reduction appears to be dependant on the base liquid, at least in the case of $Al_2O_3$.

Lastly a strong relation was observed between the sedimentation velocity and nanofluid stability. Upon comparison of theoretical and experimental results, a clear boundary was able to be demarcated between the stable and unstable nanofluids.

These novel and important findings emerged from this work are expected to provide useful guidelines to formulate stable nanofluids. More research avenues are also expected to open up to proceed along these lines.